# Spontaneous Breaking of the SU(3) Flavor Symmetry in a Quantum Hall Valley Nematic

G. Krizman[1]*, A. Kazakov[2], C.-W. Cho[3,4], V. V. Volobuev[2,5], A. Majou[1], E. Ben Achour[1], T. Wojtowicz[2], G. Bauer[6], Y. Guldner[1], B. A. Piot[3], Th. Jolicoeur[7], G. Springholz[6], L.-A. de Vaulchier[1]

[1]Laboratoire de Physique de l'Ecole normale supérieure, ENS, Université PSL, CNRS, Sorbonne Université, 24 rue Lhomond 75005 Paris, France.

[2]International Research Centre MagTop, Institute of Physics, Polish Academy of Sciences, Aleja Lotnikow 32/46, PL-02668 Warsaw, Poland.

[3]Laboratoire National des Champs Magnétiques Intenses, CNRS, LNCMI, Université Grenoble Alpes, Université Toulouse 3, INSA Toulouse, EMFL, F-38042 Grenoble, France.

[4]Department of Physics, Chungnam National University, 34134 Daejeon, Republic of Korea

[5]National Technical University "KhPI", Kyrpychova Str. 2, 61002 Kharkiv, Ukraine.

[6]Institut für Halbleiter und Festkörperphysik, Johannes Kepler Universität, Altenberger Strasse 69, 4040 Linz, Austria.

[7]Institut de Physique Théorique, Université Paris-Saclay, CNRS, CEA, 91190 Gif-sur-Yvette, France.

**Two-dimensional quantum materials can host original electronic phases that arise from the interplay of electronic correlations, symmetry and topology. In particular, the spontaneous breaking of internal symmetry that acts simultaneously on the pseudospin and the spatial degree of freedom realizes a nematic ordering. We report evidence of a quantum Hall valley nematic phase with an underlying SU(3) order parameter space obtained by a spontaneous polarization between the threefold degenerate valley pseudospins in $Pb_{1-x}Sn_xSe$ quantum wells. In the presence of a Zeeman field, we demonstrate a further control of the nematic ordering with an explicit symmetry breaking. Evidence of both spontaneous and explicit SU(3) symmetry breaking, reminiscent of the quark flavor paradigm, is of fundamental interest to shape the many body physics in a SU(3) system.**

*Contact author: gauthier.krizman@ens.fr

# I. INTRODUCTION

Quantum Hall Ferromagnets (QHFMs) [1–8] are sought-after broken-symmetry quantum Hall phases of matter in which one degree of freedom – or pseudospin (spin [9,10], valley [9,11–14], layer [15,16], …) becomes polarized. A topological QHFM can host low energy excitation states such as skyrmions [2,17–20], magnons [21–23] or charge density waves [17,24], as well as spin/charge entanglement [25]. In that sense, QHFMs offer an ideal platform to manipulate degrees of freedom and information based on spin or pseudospin for applications in spintronics, valleytronics and quantum computing [26]; and are as well of prime interest to unravel fundamental collective phenomena and spontaneous broken symmetry states. A subfamily of QHFMs that has recently drawn attention is the Quantum Hall Valley Nematic (QHVN), characterized by a broken rotational space symmetry and the appearance of preferential directions for the valley pseudospin [27,28]. In this phase of matter, not only the global symmetry that acts on the pseudospin is broken like in QHFM, but also the spatial degree of freedom.

Such a QHVN phase can be spontaneously triggered by electron-electron interactions, or induced by an external parameter that naturally splits the energy levels of the internal degree of freedom, i.e., the pseudospin. In the first case, because the kinetic electronic energy is quenched by the Landau level quantization in the QH regime, the electrons' exchange energy prevails in low disorder systems such that many-body physics becomes experimentally accessible. As a result, Coulomb interactions favor the formation of the lowest energy excitation, which corresponds to exotic quasiparticles such as pseudospin skyrmions. Additionally, in the second case, an external parameter inducing a Zeeman pseudospin splitting energy greatly enriches the QHVN phase as it can finely tune its properties (like the magnitude of the excitation gap, the domain ordering, …) [7,29]. In this work, we address both situations where the symmetry is either spontaneously or/and explicitly broken.

Up to date, QHFMs have been evidenced in several different material systems such as graphene [30], two-dimensional transition metal dichalcogenides [31], Si [13], AlAs [7,14,29] or GaAs [10,32] quantum wells showing exclusively either SU(4) or SU(2) QH ferromagnetism. While breaking of SU(3) symmetry plays a key role in theoretical models such as the quark paradigm, it has so far eluded direct experimental confirmation in condensed-matter and remains a theoretical prediction [6]. It is interesting to compare the physics of the SU(3) QH states with the quark physics. At low enough energies so that only three quarks are present, strong interactions between them have a SU(3) flavor symmetry. This global symmetry is explicitly broken by quark masses coming from the Higgs sector of the standard model. Therefore, the SU(3) flavor symmetry is exact in the so-called chiral limit of zero quark masses. In this limit, the SU(3) flavor symmetry is enlarged to a $SU(3)_L \times SU(3)_R$ chiral symmetry where SU(3) rotations are now allowed to act independently on each chiral component of the Dirac fermions. Long ago, the phenomenology of bound states of quarks has shown that this chiral symmetry is indeed spontaneously broken down to a SU(3) diagonal subgroup of the $SU(3)_L \times SU(3)_R$ parent symmetry group. This phenomenon leads to the existence of Goldstone modes that include notably the pion, the lightest bound state of quarks. The explicit symmetry breaking on top of this spontaneous breakdown consistently gives an explanation for the nonzero mass of the pion as well as its light character as a would-be Goldstone mode in the chiral limit. This interplay between explicit and spontaneous symmetry-breaking is also found in the QHVN as we demonstrate in this work.

Here, we demonstrate an SU(3) QHVN in $Pb_{1-x}Sn_xSe$ quantum wells (QW) with $x = 0.13$, where the ground Landau level degeneracy is odd, contrary to any existing QHFM observed to date. Note, however, that our results can be generalized to any Sn concentration, and even to $Pb_{1-x}Sn_xTe$ or SnTe systems which present the same SU(3) symmetry. Our work is distinguished by the singular SU(3) order parameter space created by the threefold degenerate $\bar{M}$ valleys of $Pb_{1-x}Sn_xSe$ QW in its 2D Brillouin zone (see Fig. 1(a)). We observe a strong $\nu = 3$ integer QH effect whose origin is simply the single-particle level structure and directly reflects the SU(3) degeneracy of the active valleys. Most importantly, we also observe a $\nu = 2$ QH effect which we argue is a direct manifestation of the QHVN phase in this system. In the following, we show that this nematic phase emerges clearly as a gapped QH state. In the first part of this manuscript, the nematic phase is triggered by Coulomb interactions with an associated gap indicating the formation of SU(3) valley skyrmion excitations (see Fig. 1(c)). In a second part, this QHVN phase is coupled to an in-plane magnetic field which induces a non-zero Zeeman splitting between the three $\bar{M}$ valleys (see Fig. 1(d)). The splitting of the SU(3) invariant triplet allows for a direct control over a flavor number $\gamma$ which represents each valley pseudospin.

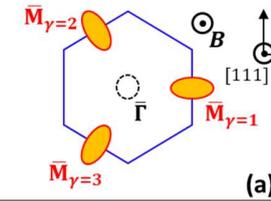

| Filling factor | $\nu = 3$ | $\nu = 3 \times 2/3$ | $\nu = 2 \times 1$ | |
|---|---|---|---|---|
| Valley population | (a) | (b) | (c) | (d) |
| Excitations | Single-particle | *Collective*: Charge density wave | *Collective*: **SU(3) Valley skyrmion** | Single-particle |
| QH gap | $\sim \hbar \omega_c$ | 0 | $\sim e^2/\varepsilon l_B$ | $\sim g^* \mu_B B_\parallel$ |
| Symmetry | $C_3$ **rotational** symmetry. Valley pseudospin $\gamma = \{1;2;3\}$ of **SU(3) order** | SU(3) flavor | **Spontaneous** Broken $C_3$ and SU(3) flavor | Broken $C_3$ and SU(3) flavor |

**FIG 1**. **Valley pseudospin polarization in the $Pb_{1-x}Sn_xSe$ system.** 2D Brillouin zone of (111)-oriented $Pb_{1-x}Sn_xSe$ QWs showing the two types of valleys, one at $\bar{\Gamma}$ and three at $\bar{M}$ identified by the flavor $\gamma$. The $\bar{M}$-valleys are equivalent in this single-particle picture obtained for instance at $\nu = 3$ (**a**). Three different situations can occur at $\nu = 2$. When $\nu = 3 \times 2/3$, the electrons are delocalized in the three valleys and form a (gapless) charge density wave where SU(3) symmetry is preserved (**b**). A QHVN phase emerges with a non-zero QH gap when two valley-pseudospins become polarized ($\nu = 2 \times 1$), triggered by collective effects (**c**) or by an in-plane Zeeman field (**d**). Depending on the mechanism, the QH gap magnitude is dominated by different energy scales. The Coulomb effects induce SU(3) valley skyrmions while the Zeeman field allows for a control over the flavor number $\gamma$.

## II. RESULTS AND DISCUSSIONS

### A. Observation of the spontaneous QHVN ordering

Magneto-transport on a 23 nm-thick $Pb_{0.87}Sn_{0.13}Se$ QW has been measured under perpendicular magnetic field up to $B = 35$ T at $T = 100$ mK. The resulting longitudinal resistance curve is plotted in Fig. 2(a). As demonstrated in our previous work on integer QH effect in $Pb_{1-x}Sn_xSe$ QWs [33,34], the observed filling factor series verifies the Dirac expression $\nu = 2g_{v,\bar{M}}(n + 1/2)$ with $g_{v,\bar{M}} = 3$ being the valley degeneracy (the factor 2 stands for the spin degeneracy). This probes that, in this sample, the in-plane biaxial strain induced by the lattice mismatch with the Pb$_{0.9}$Eu$_{0.1}$Se buffer is such that the threefold degenerate $\bar{M}$ valleys are shifted to higher energy (~50 meV) with respect to the single $\bar{\Gamma}$ valley. By means of a proper adjustment of the Fermi level, this allows the configuration where only the three $\bar{M}$ valleys are populated with hole carriers and the $\bar{\Gamma}$ valley is completely empty. Indeed, the most pronounced $R_{xx}$ minima are seen at $\nu = 15$, $\nu = 9$ and $\nu = 3$ indicating when the Fermi energy lies either between the Landau levels $n = 3$ and $n = 2$; or between $n = 2$ and $n = 1$; or between $n = 1$ and $n = 0$ respectively. A smaller feature is seen for $\nu = 6$. To account for these observations, we have modeled the Landau levels of this anisotropic multivalley Dirac compound within the $\mathbf{k} \cdot \mathbf{p}$ framework [35–37] (see Fig. 2(b) and supplementary materials [38]) using band parameters determined by magneto-optics and magneto-transport measurements. Such Landau levels are depicting 2D massive Dirac fermions with a small 2$^{nd}$ order correction [35,36,39] responsible for the slight spin splitting between $n\uparrow$ and $n\downarrow$ Landau levels.

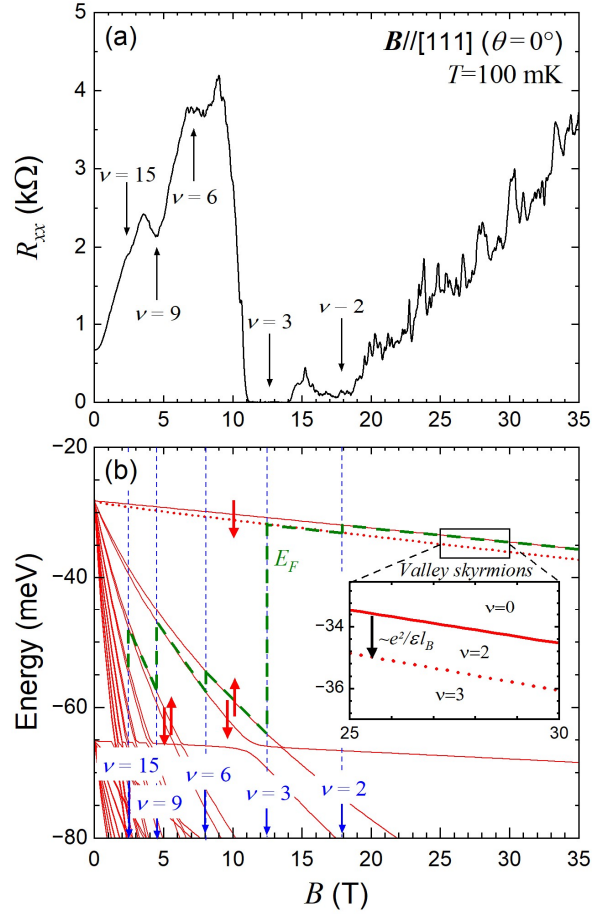

**FIG 2**. Spontaneous QHVN phase and SU(3) valley skyrmion. (a) Longitudinal resistance $R_{xx}$ at $T = 100$ mK as a function of the perpendicular magnetic field. The minima correspond to integer filling factors that are indicated. (b) Landau levels calculated using the band parameters given in the supplementary materials. The red dotted line indicates the emerging gap due to a SU(3) valley skyrmion formation and is calculated following $E_{n=0}(B) - e^2/\varepsilon l_B$ where $E_{n=0}(B)$ is the magnetic field dispersion of the ground Landau level. The inset shows a zoom-in of the valley skyrmion excitation gap. The Landau level are threefold valley degenerate, and their spins are indicated by vertical red arrows. Note that the $n = 0$ Landau level is spin-polarized. The filling factors corresponding to the observed $R_{xx}$ minima are indicated in blue, and the Fermi level evolution is indicated in green.

The measured QH effect depicted in Fig. 2(a) is well-reproduced by the model. The wide $R_{xx}$ zero-resistance state observed at $\nu = 3$ around $B = 12$ T corresponds to the large cyclotron gap between the $n = 0$ and $n = 1$ Landau levels (~30 meV here). Due to the Dirac nature of the Pb$_{1-x}$Sn$_x$Se system [40], the $n = 0$ Landau level is spin-polarized so its degeneracy involves only the valley degree of freedom. The $\nu = 3$ state has thus a single-particle origin since it appears when there is complete filling of the three $n = 0$ Landau levels. As a result, the threefold degeneracy with SU(3) symmetry is naturally obtained in the lead salt compounds. Additionally, the small minimum observed at $\nu = 6$ is attributed to the small Zeeman spin splitting calculated between the Landau levels $n = 1\uparrow$ and $n = 1\downarrow$. Note that under perpendicular magnetic field, the Landau levels emerging at the $\bar{M}$ valleys are invariant with respect to the $C_3$ rotational symmetry, and thus, each Landau level is threefold valley degenerate in a single particle picture.

Most importantly, we also observe a weaker $R_{xx}$ minimum at $\nu = 2$ around $B = 18$ T in Fig. 2(a), which indicates the presence of an excitation gap which is smaller than the one corresponding to $\nu = 3$. This is direct evidence of the spontaneous lifting of the threefold valley degeneracy and thus, the emergence of the valley-polarized QHFM ordering, i.e., the QHVN phase. In this phase, one $\bar{M}$ valley is electronically emptied while the two others remain filled and are responsible for the appearance of the resistance minimum at filling factor $\nu = 2$ as schematically illustrated in Fig. 1(c). The lowest energy neutral excitations in this situation are expected to be valley skyrmions. Indeed, if the spin part is fully symmetric with respect to electron exchange, one has to build a fully antisymmetric spatial wavefunction with maximal avoidance between the charge carriers. This realizes the preferred minimal energy state when there is only the Coulomb interaction between the carriers and no kinetic energy involved e.g., the case of Landau levels in a 2D system. The excitations are then highly collective in nature and involve a smooth pseudospin texture called "skyrmion" [41]. In such a state the pseudospin flip is delocalized over a spatial scale that diverges when the Landau levels associated to each pseudospin are degenerate in the single particle picture. The skyrmion excitation gap is proportional to the Coulomb energy scale $e^2/\varepsilon l_B \sim (56\sqrt{B})/\varepsilon$ [meV], where $l_B = \sqrt{\hbar/eB}$ denotes the magnetic length and $\varepsilon$ the dielectric constant.

The spontaneously opened QHVN gap observed at $\nu = 2$ is expected to be of the order of 1.2 meV in our case as it appears around $B = 18$ T (see the dotted line in Fig. 2(b)). Despite the rather high static dielectric constant of the PbSe material of ~200 [42,43], our observation is a clue that Coulomb interaction effects are present at temperatures in the 100 mK region. While such valley skyrmions have been convincingly observed in the AlAs system [18] with underlying SU(2) symmetry, here, we find traces of these topological entities in the case of a SU(3) pseudospin symmetry. Since the skyrmion texture is a smooth pseudospin flip, it is straightforward to embed a two-state SU(2) skyrmion into any larger number of components $N > 2$ under a SU($N$) order parameter space. Already long ago it was proposed that this exhausts the excited states of the system with any number of degenerate components [44]. While this theoretical construction shows how to generalize the known skyrmions to any number of flavors, this did not rule out the existence of other exotic excited states lower in energy. Numerical investigations in the SU(4) paradigm, aiming at the case of graphene, have found no evidence for any state beyond the standard skyrmions [20].

## B. QHVN phase promoted by in-plane field

Another strategy to induce a SU(3) QHVN phase in this system is to use an external parameter acting selectively on the valley pseudospin. For this purpose, the magnetic field has been tilted towards the in-plane $[\bar{1}\bar{1}2]$-direction, parallel to the long axis of one of the three anisotropic $\bar{\text{M}}$ valleys, as illustrated in Fig. 1(d). As a result, the flavor $\gamma = 1$ which is along $B_\parallel$ is intuitively expected to behave differently compared to the flavors $\gamma = 2$ and $\gamma = 3$ which remain equivalent. In this configuration where a strong Zeeman energy is induced between the valley pseudospins, we cross over to a situation in which full pseudospin polarization is not due to Coulomb interactions but to complete occupancy of the one-particle polarized states [41]. In this limit, we expect an energy gap corresponding to a single electron pseudospin flip given by an in-plane Zeeman-like energy $\sim g^* \mu_B B_\parallel$ with $g^*$ an effective $g$-factor [6].

The measured $R_{xx}$ and $R_{xy}$ resistance curves at $T = 300$ mK are plotted in Fig. 3(a-f) for different tilt angles $\theta$ of the magnetic field with respect to the [111]-direction. Under tilted magnetic field, the series of the observed QH states changes drastically compared to the $\theta = 0°$ case depicted in Fig. 2. Instead, with increasing tilt angle, we observe a vanishing $\nu = 3$ QH plateau that completely disappears for $\theta \geq 60°$, and, oppositely, the $\nu = 2$ plateau is getting more and more pronounced as witnessed by the $R_{xy}$ curves at $\theta = 50°$ and $\theta = 60°$. The strengthened $\nu = 2$ plateau is the direct demonstration of the larger lifting of the $\bar{\text{M}}$ valleys degeneracy upon tilted magnetic field, namely, the explicit breaking of the SU(3) symmetry and the emergence of the Zeeman-induced QHVN phase.

This behavior is well-reproduced by the Landau levels shown in Fig. 3(g-i) calculated within a single-particle model for the same band parameters as in Fig. 2 [38]. As the tilt angle increases, the QH gap corresponding to $\nu = 3$, $\Delta^{\nu=3}$, decreases and eventually vanishes for $\theta \geq 60°$ (see Fig. 3(h)), while the QH gap $\Delta^{\nu=2}$ between the ground Landau levels of the $\gamma = 1$ flavor with respect to the degenerate flavors $\gamma = 2$ and $\gamma = 3$ clearly increases, as witnessed by the pronounced plateau seen for $\theta = 60°$ (see Fig. 3(h)). Because the Fermi surfaces of the three $\bar{\text{M}}$-valleys are elliptical and not spherical due to the anisotropy in the $Pb_{1-x}Sn_xSe$ compounds [45,46], the Landau levels are $\gamma$-dependent against an in-plane component of the magnetic field. Consequently, the Zeeman valley splitting, i.e., the valley degeneracy, can be manipulated with an in-plane magnetic field within the single particle picture. This anisotropy [38] is then directly responsible for the explicit breaking of the SU(3) flavor symmetry and the emergence of the $\nu = 2$ QH plateau when a non-zero in-plane magnetic field component is applied. Ultimately, the anisotropy is adjusted in such a way that the calculations give a remarkable qualitative agreement with the observations. This allows us to determine the QH gaps corresponding to several filling factors, as represented in Fig. 3(j), where their evolutions are shown as a function of the tilt angle.

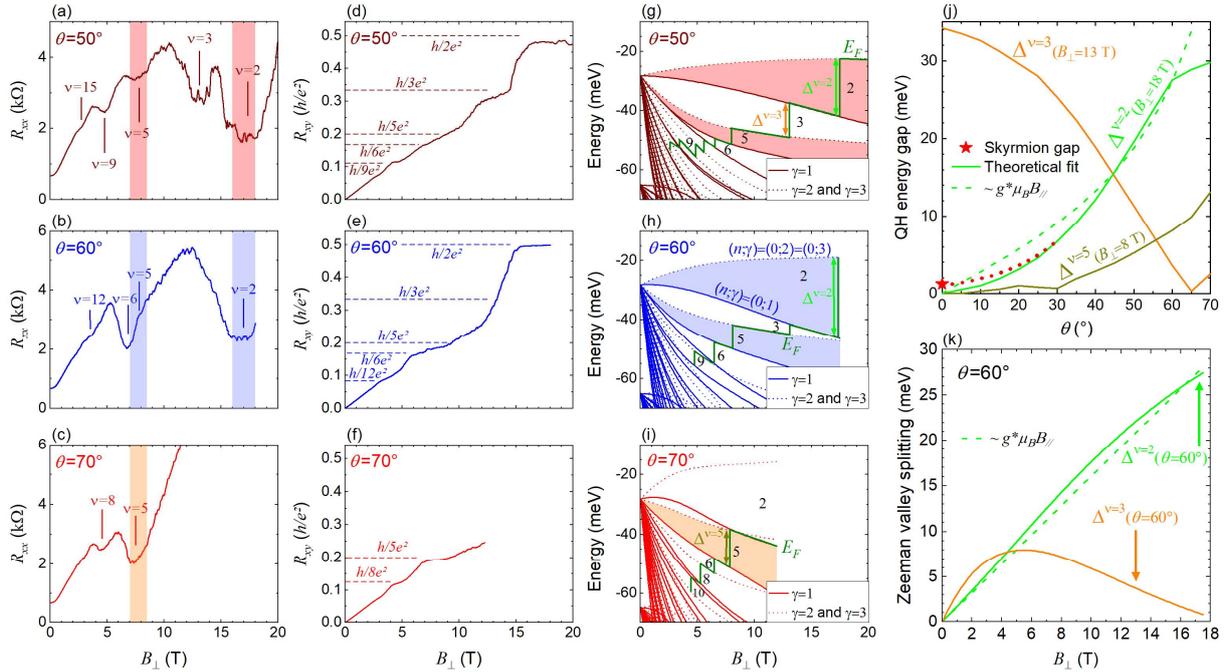

**FIG 3. QHNV phase controlled by in-plane magnetic field.** QH measurements showing $R_{xx}$ **(a-c)** and $R_{xy}$ **(d-f)** at $T = 300$ mK and for tilted magnetic fields. The shaded areas indicate the vicinity of $\nu = 2$ and $\nu = 5$ that stand as direct evidence of the QHVN phase. **(g-i)** Calculated Landau levels using the parameters given in the supplementary materials at $\theta = 50°$, $\theta = 60°$ and $\theta = 70°$. The Fermi level is indicated, as well as the corresponding filling factors in between Landau levels. The solid (dashed) lines correspond to the Landau levels emerging at the $\overline{M}_{\gamma=1}$ valley ($\overline{M}_{\gamma=2}$ and $\overline{M}_{\gamma=3}$ valleys). The QH gaps at $\nu = 2$, $\nu = 3$ and $\nu = 5$, occurring at $B_\perp = 18$ T, $B_\perp = 13$ T and $B_\perp = 8$ T respectively, are indicated by an arrow and denoted as $\Delta^\nu$. **(j)** QH gaps $\Delta^\nu$ versus tilt angle. The skyrmion gap is indicated and schematically smoothly linked by a red dotted line to the QH gap due to the Zeeman valley field. **(k)** Zeeman valley splitting giving the filling factors $\nu = 2$ (green solid line) and $\nu = 3$ (orange) versus the perpendicular-component of the magnetic field at $\theta = 60°$. In panels (j) and (k) the green dashed line is calculated using a $g^*\mu_B B_\parallel$ dependence taking $g^* = 16$.

The $\nu = 2$ QH gap $\Delta^{\nu=2}$ calculated within our $\boldsymbol{k}.\boldsymbol{p}$ model can be fitted using an in-plane Zeeman law scaling as $\sim g^*\mu_B B_\parallel$, which has been used in Fig. 3(j,k) in dashed lines versus tilt angle and magnetic field. A great agreement is found for an effective $g$-factor $g^* = 16$. This demonstrates that, at high tilt angles, the nematic state is governed by the in-plane magnetic field that acts as a valley Zeeman field in the single particle picture. For lower tilt angles, the Zeeman valley splitting goes to zero so that it becomes negligible with respect to the Coulomb interaction, as shown in Fig. 3(j). However, a nonzero gap is still observed as witnessed by the data in Fig. 2, meaning that the QH gap changes its nature to collective excitations. The crossover between these two regimes is characterized by a competition between Coulomb interaction and Zeeman field [18,47,48].

Furthermore, we observe two emerging QH plateaus $\nu = 5$ and $\nu = 8$ as the tilt angle increases. These plateaus demonstrate the manifestation of a nematic valley ordering also for the higher index Landau levels $n = 1$ and $n = 2$. Indeed, these filling factors are obtained by lifting the degeneracy of the $n = 1$ and $n = 2$ Landau levels of the $\overline{M}$ valleys. At $\nu = 5$ for instance, the $(n; \gamma) = (1; 1)$ Landau level is empty while the two $(1; 2) = (1; 3)$ are filled, as shown in Fig. 3(i) for $\theta = 70°$. Our theoretical calculations give a sizeable energy gap at high tilt angles ($\Delta^{\nu=5} \sim 10$ meV) for this higher index Landau levels QHVN phase (see Fig. 3(j)), which gives a good agreement with the observations.

## III. CONCLUSIONS

In summary, we have observed a spontaneous polarization of the SU(3) flavor in the $Pb_{1-x}Sn_xSe$ 2D system reflected by the appearance of a $\nu = 2$ plateau under perpendicular magnetic field. This clearly indicates that when the three valleys are at a filling factor $\nu = 2/3$, the QH system prefers being in a polarized state where two valleys are at $\nu = 1$ and one single valley is at $\nu = 0$, giving a net filling factor of $\nu = 2$ with a gapped QH state associated with a SU(3) valley skyrmion excitation. In addition, the QHVN phase is also evidenced and explicitly induced by an in-plane magnetic field responsible for a Zeeman valley splitting. This is an example of a valley nematic phase with an underlying SU(3) order parameter space, reminiscent of the quark physics.

These observations motivate further experimental and theoretical work to obtain additional information on the transport anisotropy, the coherence length and the non-local transport in this particular QHVN phase. Furthermore, it opens new perspectives for the demonstration of other collective electronic phases in a unique SU(3) QH system such as the QH ferroelectricity [49,50] predicted to occur in lead salt compounds. Fractional quantum Hall states are also expected to be substantially enriched by the SU(3) symmetry. We certainly expect to find many states obtained by a direct generalization of the composite fermion construction. Interestingly, exact diagonalizations of SU(3) systems have given hints of physics for $\nu = 2/3$ beyond the composite fermion paradigm [51].

## ACKNOWLEDGMENTS

The authors warmly thank M. Shayegan for enlightening discussions. We thank the financial support from the Agence Nationale de la Recherche (grant No ANR-24-CE91-0014-01) and the Austria Science Funds (project PIN6540324). We thank the support from LNCMI-CNRS, member of the European Magnetic Field Laboratory (EMFL). V.V.V. also acknowledges the long-term program of support for Ukrainian research teams at the Polish Academy of Sciences, carried out in collaboration with the U.S. National Academy of Sciences and funded by external partners. This research was partially supported by the "MagTop" project (FENG.02.01-IP.05-0028/23) carried out within the "International Research Agendas" programme of the Foundation for Polish Science cofinanced by the European Union under the European Funds for Smart Economy 2021-2027 (FENG) and by Narodowe Centrum Nauki (NCN, National Science Centre, Poland) IMPRESS-U Project No. 2023/05/Y/ST3/00191.

# Supplementary Materials for

# Spontaneous Breaking of the SU(3) Flavor Symmetry in a Quantum Hall Valley Nematic

G. Krizman, A. Kazakov, C.-W. Cho, V. V. Volobuev, A. Majou, E. Ben Achour, T. Wojtowicz, G. Bauer, Y. Guldner, B. A. Piot, Th. Jolicoeur, G. Springholz, L.-A. de Vaulchier

### I. SAMPLES FABRICATION AND PARAMETERS

The characteristics of the samples investigated in this work are listed in Table S1. Each quantum well (QW) has a thickness of 23 nm and is buried between $Pb_{0.9}Eu_{0.1}Se$ 1 µm buffer and 150 nm cap semi-insulating layers. The samples are grown by molecular beam epitaxy onto (111)-oriented $BaF_2$ substrates.

The contacts are made by indium alloying. The Hall bar of Sample 1 (in the main text) is 22x5 µm² large while Sample 2 (in the Supplementary materials) is 1000x275 µm².

The parameter $\Delta_{\bar{\Gamma}-\bar{M}}$ denotes the energy splitting between the single $\bar{\Gamma}$ and three $\bar{M}$ valleys and is further defined in Ref. [1]. The anisotropy of the constant energy contour of each valley is taken into account by defining the transverse and longitudinal electron velocities $v_t$ and $v_l$ respectively, as well as the transverse and longitudinal $g$-factors $g_t$ and $g_l$ respectively (see the next subheading). These quantities are both defined in the local axis of the ellipsoids. The in-plane biaxial strain $\varepsilon_\parallel$ has been measured by X-ray diffraction; the band structure (the energy gap $2\delta$ and the electron velocities) has been characterized by magneto-optics; the others features have been deduced from magneto-transport experiments (see further Sections and main text).

**Table S1.** Samples characteristics.

| Samples | Sample 1 | Sample 2 |
|---|---|---|
| $d$ [nm] | 23±1 | 23±1 |
| $x_{Sn}$ [%] | 13±1 | 11±1 |
| $\varepsilon_\parallel$ [%] | 0.39±0.04 | 0.45±0.04 |
| $\Delta_{\bar{\Gamma}-\bar{M}}$ [meV] | 55 | 65 |
| $2\delta$ [meV] | 55±2.5 | 70±2.5 |
| Doping [cm$^{-2}$] | p=8.9x10$^{11}$ | p=1.3x10$^{12}$ |
| Mobility [cm²/V.s] | $\mu$ = 50 000±2000 | $\mu$ = 30 000±2000 |
| $v_t$ [x10$^5$ m/s] | 5.00±0.05 | 5.10±0.05 |
| $v_l$ [x10$^5$ m/s] | 4.40±0.05 | 4.40±0.05 |
| $g_t$ | 7±5 | 7±4 |
| $g_l$ | 34±5 | 18±3 |

## II. $\boldsymbol{k}\cdot\boldsymbol{p}$ DERIVATION OF ANISOTROPIC MULTIVALLEY DIRAC LANDAU LEVELS

The procedure for computing the Landau levels of the PbSnSe multivalley 2D Dirac material has been given in Ref. [2] in the case of a magnetic field perpendicular to the (111) surface. This gives two sets of Landau levels, one for the electrons at the $\bar{\Gamma}$-point, and the other, threefold degenerate, corresponding to the $\bar{M}$-point of the 2D Brillouin zone. The calculation process is the following:

The $\boldsymbol{k}\cdot\boldsymbol{p}$ Hamiltonian needs to be expressed in the sample coordinates $(x, y, z)$ with $z//[111]$ perpendicular to the surface. Thus, the valley local axes of the $\bar{M}$-point are rotated along the $y$-axis by an angle $\omega = 70.53°$ ($\cos\omega = 1/3$ and $\sin\omega = 2\sqrt{2}/3$). After solving exactly the quantum well problem at $k_x = k_y = 0$ and finding the eigenenergies and envelope wavefunctions of the confined states, the $k_x, k_y$-terms (or after Peierls substitution, the $B$-terms) are treated in perturbation. To do so, the perturbative terms are written in a basis of envelope wavefunctions obtained at $k_x = k_y = 0$. The resulting Hamiltonian is then diagonalized to find the eigenenergies corresponding to the $k_x, k_y$-dispersion (or Landau levels). In our case, the tilted magnetic field implies additional terms to be treated as perturbations compared to those in Ref. [2]. They are detailed in the following with the band parameters defined in Ref. [2].

For a tilted magnetic field by an angle $\theta$ with respect to the [111]-direction, written as

$$\boldsymbol{B} = \left(B\sin\theta\cos(\alpha_\gamma - \phi)\,;\, B\sin\theta\sin(\alpha_\gamma - \phi)\,;\, B\cos\theta\right)$$
$$= \left(B_\parallel \cos(\alpha_\gamma - \phi)\,;\, B_\parallel \sin(\alpha_\gamma - \phi)\,;\, B_\perp\right)$$

a suitable choice of gauge for the vector potential is:

$$\boldsymbol{A} = \begin{bmatrix} B_\parallel z \sin(\alpha_\gamma - \phi) \\ B_\perp x - B_\parallel z \cos(\alpha_\gamma - \phi) \\ 0 \end{bmatrix}$$

Here, $(x, y, z)$ are the coordinates of the sample with $z//[111]$, $\alpha_\gamma$ are the azimuthal angle of the three $\bar{M}$-valleys that are for $\gamma = \{1,2,3\}$, $\alpha_\gamma = \{0, 2\pi/3, -2\pi/3\}$ respectively. $\phi$ is the azimuthal angle of the magnetic field with the valley orientation. In our geometry, $\phi = 0$ holds, as depicted in Fig. S1.

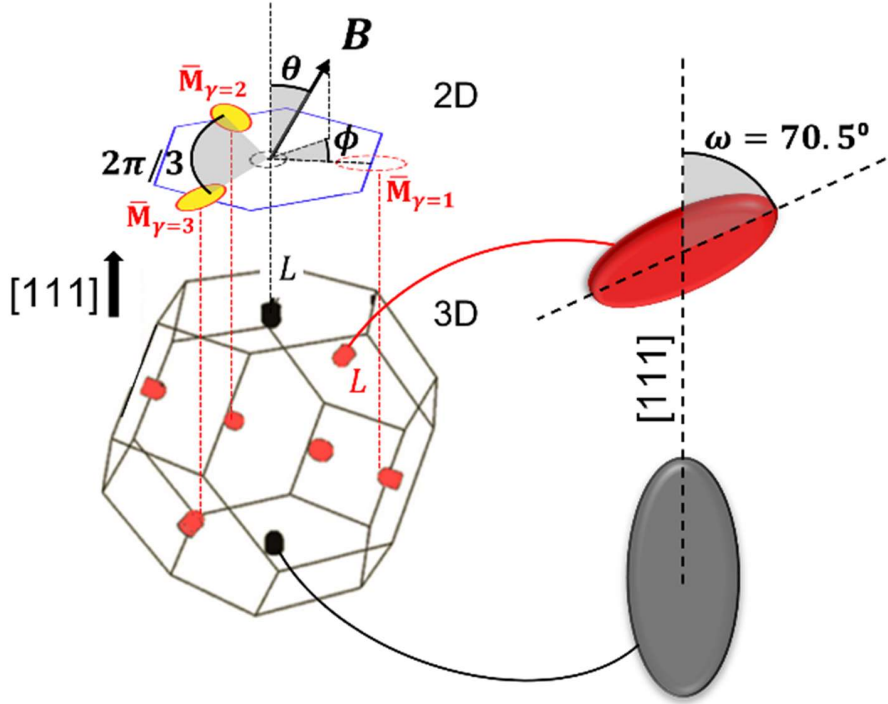

**FIG S1.** Schematic diagram of the 3D and 2D Brillouin zones of the $Pb_{1-x}Sn_xSe$ cubic system. In 3D, the four electron pockets appearing at the $L$-points are indicated. They are ellipsoids with major axis oriented along [111] (in black) or tilted by an angle $\omega$ with respect to [111] (in red). These valleys, projected onto the (111)-plane, give the electron (or hole) pockets at the $\bar{\Gamma}$ and the $\bar{M}$ points respectively in the 2D representation. The magnetic field makes an angle $\theta$ with the [111]-direction, and an angle $\phi$ with the $[\bar{1}\bar{1}2]$-direction. The angle between two $\bar{M}$-valleys is $2\pi/3$. In our geometry, $\phi = 0$.

Together with the Hamiltonian developed in Ref. [2], the following term has to be added to the treatment in perturbation:

$$\delta H(B) = \begin{pmatrix} 0 & 0 & c & a \\ 0 & 0 & a^* & -c \\ c & a & 0 & 0 \\ a^* & -c & 0 & 0 \end{pmatrix}$$

with $a = eB_{\parallel}z[iv_t e^{-i(\alpha_\gamma - \phi)} - \sin(\alpha_\gamma - \phi)\sin^2\theta\,(v_t - v_l)]$
and $c = -eB_{\parallel}z\sin(\alpha_\gamma - \phi)\sin\theta\cos\theta\,(v_t - v_l)$.
$v_t$ and $v_l$ are respectively the transverse and longitudinal velocities in the frame of the valley ellipsoid.

Furthermore, a Zeeman term has to be added and write $H_z = \frac{1}{2}\mu_B \boldsymbol{B}\bar{\bar{g}}\boldsymbol{S}$ with $\bar{\bar{g}}$ being the $g$-tensor

$$\bar{\bar{g}} = \begin{pmatrix} g_t & 0 & 0 \\ 0 & g_t & 0 \\ 0 & 0 & g_l \end{pmatrix}$$

in the frame of the ellipsoid, and after rotation:

$$S = \begin{pmatrix} -\sin\theta\, \sigma_z + \cos\theta\, (\cos\alpha_\gamma\, \sigma_x + \sin\alpha_\gamma\, \sigma_y) \\ -\sin\alpha_\gamma\, \sigma_x + \cos\alpha_\gamma\, \sigma_y \\ \cos\theta\, \sigma_z + \sin\theta\, (\cos\alpha_\gamma\, \sigma_x + \sin\alpha_\gamma\, \sigma_y) \end{pmatrix} \quad \text{and} \quad B = \begin{pmatrix} \frac{B_\parallel}{3}\cos(\alpha_\gamma - \phi) - \frac{2\sqrt{2}}{3}B_\perp \\ -B_\parallel \sin(\alpha_\gamma - \phi) \\ \frac{B_\perp}{3} + \frac{2\sqrt{2}}{3}B_\parallel \cos(\alpha_\gamma - \phi) \end{pmatrix}$$

After some calculations, one obtains the Zeeman term to be treated in perturbation as:

$$H_z(B) = \frac{1}{2}\mu_B \begin{pmatrix} G_z & G_\parallel & 0 & 0 \\ G_\parallel^* & -G_z & 0 & 0 \\ 0 & 0 & -G_z & -G_\parallel \\ 0 & 0 & -G_\parallel^* & G_z \end{pmatrix}$$

with

$$\begin{cases} G_\parallel = \dfrac{2\sqrt{2}}{9}B_\perp(g_l - g_t)e^{-i\alpha_\gamma} + \dfrac{8}{9}B_\parallel g_l e^{-i\alpha_\gamma}\cos(\alpha_\gamma - \phi) \\ \quad + B_\parallel g_t\left[\dfrac{1}{9}\cos(\alpha_\gamma - \phi)\,e^{-i\alpha_\gamma} + i\sin(\alpha_\gamma - \phi)\,e^{-i\alpha_\gamma}\right] \\ G_z = \dfrac{B_\perp}{9}(g_l + 8g_t) + \dfrac{2\sqrt{2}}{9}(g_l - g_t)B_\parallel \cos(\alpha_\gamma - \phi) \end{cases}$$

These two terms $\delta H(B)$ and $H_z(B)$ are added to the treatment detailed in Ref. [2] and the Landau levels are then solved and found numerically. In this model, the angle $\alpha_\gamma$ discriminates the Landau levels emerging at different $\bar{\text{M}}$-points. With $\phi = 0$, $\gamma = 2$ and $\gamma = 3$ are equivalent but $\gamma = 1$ behaves differently as long as the system stays anisotropic. This model describes accurately the Landau levels of a 2D Dirac anisotropic multivalley material under tilted magnetic field.

## III. EXPERIMENTAL DETERMINATION OF THE ANISOTROPIC $g$-FACTORS IN PBSNSE QUANTUM WELLS

In order to accurately determine the Landau levels structure of $Pb_{1-x}Sn_xSe$ QWs and in particular, the anisotropic $g$-factors, we measured the magneto-transport properties of another sample (Sample 2) whose properties are very similar to the sample in the main text (Sample 1), listed in Table S1. The measurements have been performed between $T = 1.2$ K and $T = 25$ K to determine the magnitude of large energy gaps via the temperature-dependent resistivity, and up to $B = 35$ T. The Fermi energy lies in the valence band of the $\overline{M}$ valleys only, thus, our experiments here probe only the holes properties of a $Pb_{1-x}Sn_xSe$ QW with x=11 %.

Figure S2 shows the temperature dependence of the longitudinal and transverse resistances $R_{xx}$ and $R_{xy}$ at two different tilted angles $\theta = 0°$ and $\theta = 55°$. At $\theta = 0°$, the observed QH plateaus follow the sequence $\nu = 6(n + 1/2)$ similar as for the sample in the main text (Sample 1), meaning that only the three $\overline{M}$ valleys are populated. Only a faint $R_{xx}$ minimum is seen at $\nu = 6$ for $T = 1.2$ K and $T = 5$ K, showing that the Zeeman spin splitting between the two Landau levels $n = 1$ is rather small at $\theta = 0°$, in perfect agreement with the observation discussed in the main text.

At high tilt angles such as $\theta = 55°$, the plateau $\nu = 6$ clearly emerges while the $\nu = 9$ plateau vanishes, as seen in Fig. S2(c,d). Only a small feature appears in the $R_{xx}$ curve at $\nu = 9$, which indicates that the gap is almost closed. This demonstrates that a tilted magnetic field significantly strengthens the Zeeman spin splitting between the two $n = 1$ Landau levels in this system. Moreover, the QH plateau at $\nu = 3$ is still observed but the $R_{xx}$ minima are less pronounced than for the $\theta = 0°$ case. This indicates a significantly smaller QH gap corresponding to this plateau.

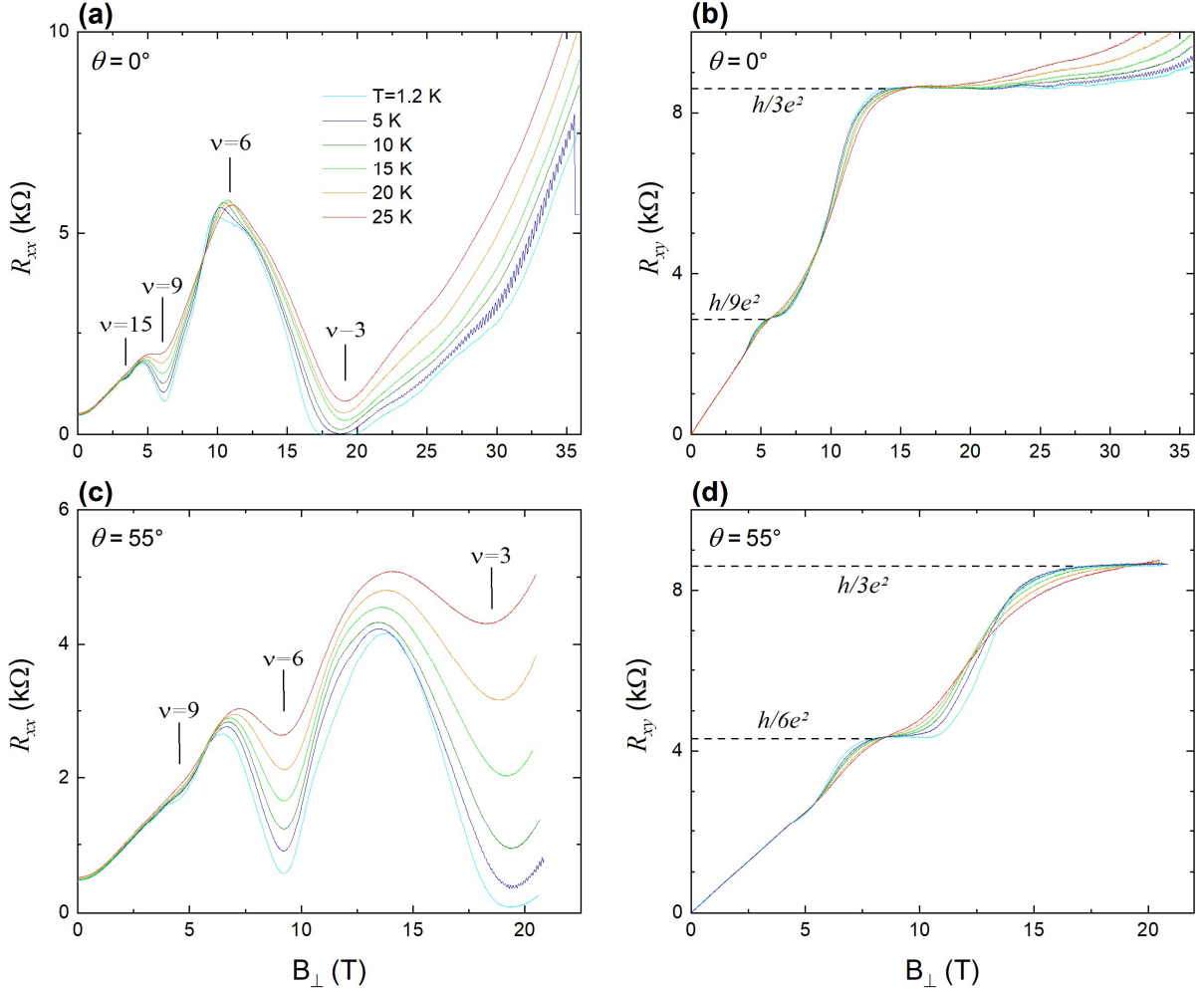

**FIG S2.** (**a,b**) Measurements of the $R_{xx}$ (a) and $R_{xy}$ (b) resistances of Sample 2 up to $B = 35$ T for a perpendicular magnetic field ($\theta = 0°$) and for different temperatures. The $R_{xx}$ minima are indicated as well as the quantized $R_{xy}$ plateaus. (**c,d**) Similar as (a,b) under tilted magnetic field with $\theta = 55°$.

In order to address this Zeeman splitting more quantitatively and to confront our model developed in Section II, we used the Lifshitz-Kosevich (LK) formula to determine the QH gap corresponding to the different plateaus observed at $\theta = 0°$ and $\theta = 55°$. The LK formula writes

$$\Delta R_{xx}/R_{max} \sim X/\sinh X \quad \text{with} \quad X = 2\pi^2 k_B T/\Delta$$

The fitting of the experimental data $\Delta R_{xx}/R_{max}$ with this formula gives the satisfactory results plotted in Fig. S3. In this picture, the gap Δ corresponds to the energetic separation between the centers of two Landau levels. For the $\nu = 3$ QH plateau for instance, an energy gap of Δ= 39 meV is found and gives precisely the well-known cyclotron effective mass in Pb$_{1-x}$Sn$_x$Se QW [3–5].

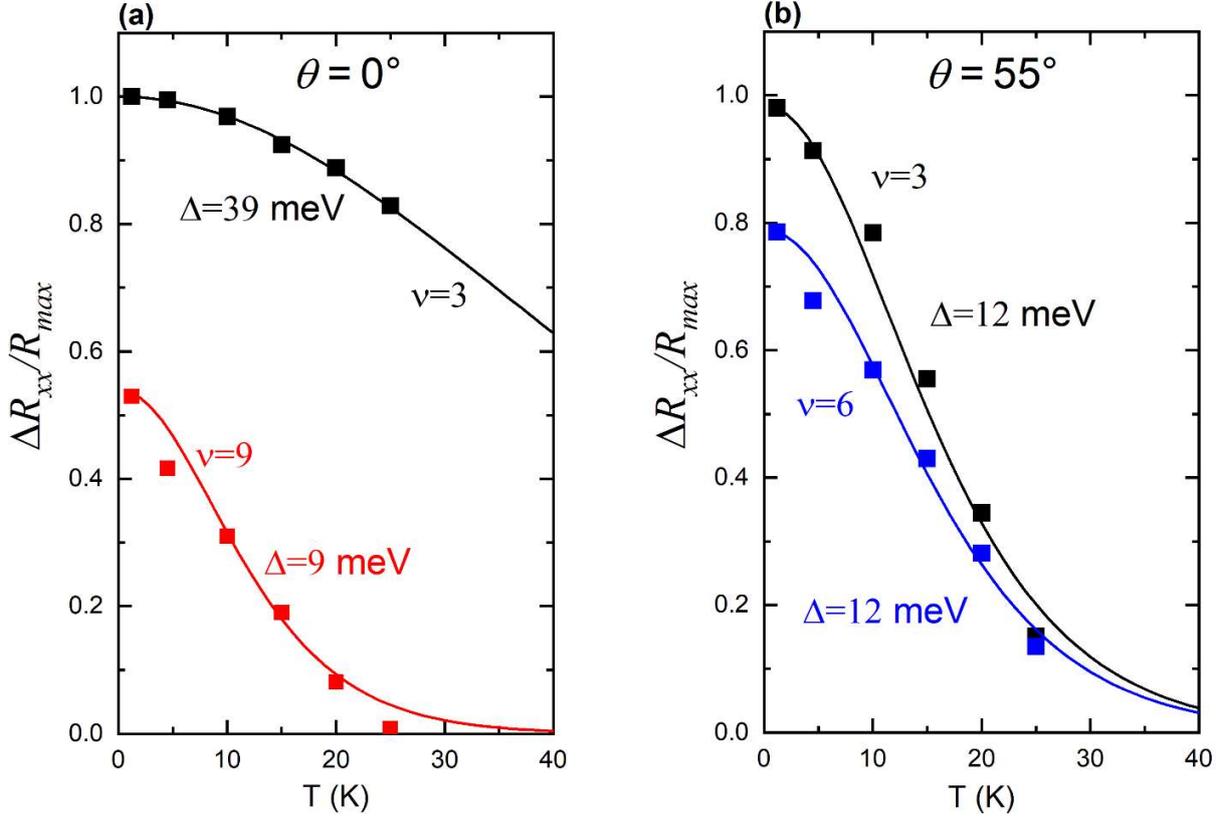

**FIG S3.** $\Delta R_{xx}/R_{max} = (R_{max} - R_{min})/R_{max}$ quantities of Sample 2 extracted from the measurements presented in Fig. S2 at $\theta = 0°$ (**a**) and $\theta = 55°$ (**b**) at different filling factors. The solid lines denote the best fits to the experimental data using the LK formula.

Figure S4 compares the experimental results with the theory. The Landau levels of the three $\overline{M}$ valleys are calculated at $\theta = 0°$ and $\theta = 55°$ and the experimental QH gaps obtained in Fig. S3 are indicated by vertical segment lines. The best fit is obtained for the parameters listed in Table S1. $g_t$ and $g_l$ being the only free parameters for this fit, they are accurately determined as $g_t = 7$ and $g_l = 18$ in this sample. Taking these $g$-factors, theory and experiments match perfectly. Under tilted magnetic field, the gap of the $\nu = 3$ plateau shrinks to $\Delta = 12$ meV which is in great agreement with the theory. The emergence of the $\nu = 6$ plateau is well-explained by the increase of the Zeeman splitting for $\theta > 0°$.

Interestingly, the theory gives the presence of a small QH gap (~3 meV) for $\nu = 5$, corresponding to the lift of degeneracy of the $n = 1$ Landau levels of the two types of $\overline{M}$-valleys under tilted magnetic field and the appearance of the high order Landau levels QHVN phase, as observed in the main text for the other sample as well. This nematic ordering is responsible for the very slight kink observed in the $R_{xx}$ curve at $B_\perp = 11$ T for $T = 1.2$ K (see Fig. S2(c)).

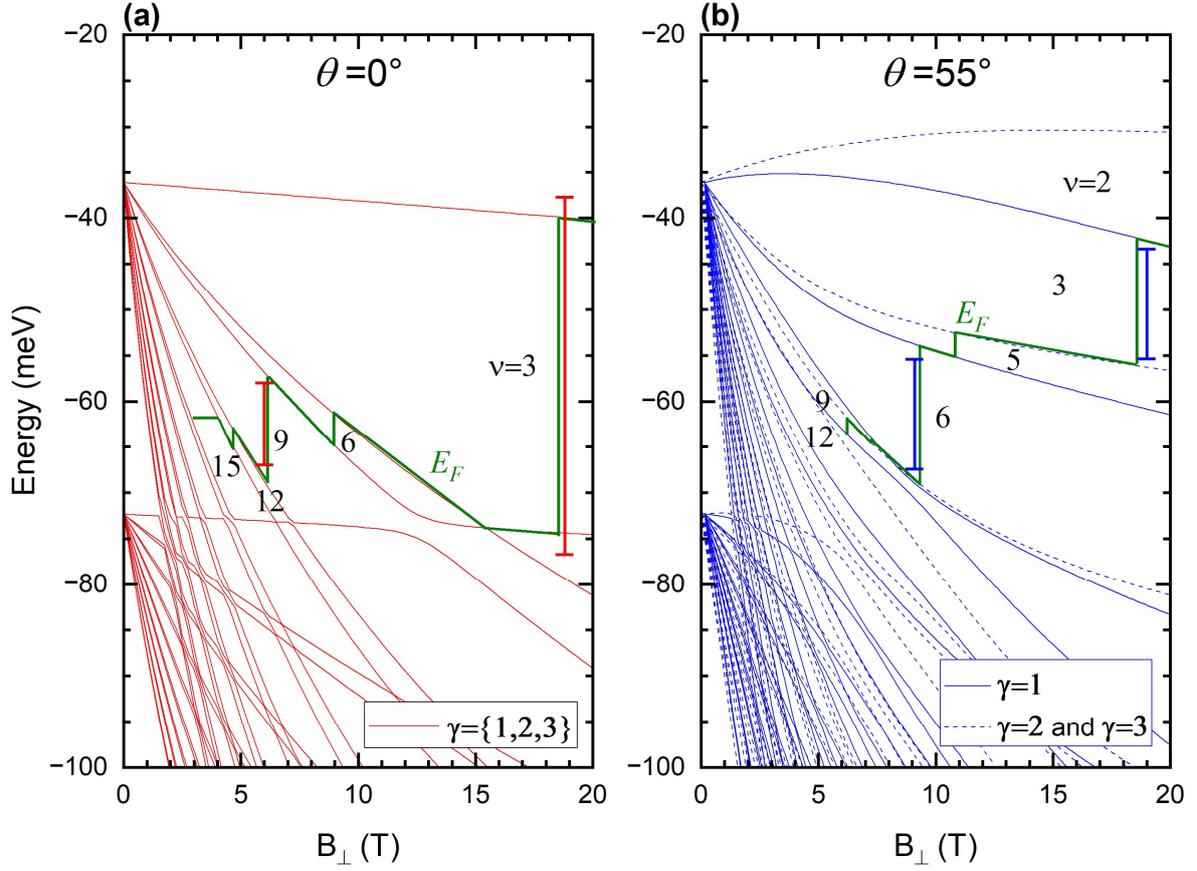

**FIG S4.** Calculated Landau levels of the three different flavors of the $\bar{M}$-valleys using the model developed in Section II for Sample 2 at $\theta = 0°$ (**a**) and $\theta = 55°$ (**b**). The parameters used are listed in Table S1. The Fermi energy is indicated in a green solid line, and the filling factors are written as numbers in between Landau levels. The experimentally determined QH gaps are indicated by vertical red and blue line segments corresponding to the plateaus observed at $\theta = 0°$ and $\theta = 55°$ respectively.

## IV. THE $R_{xy}$ MEASUREMENT OF SAMPLE 1 AT T=100 mK AND $\theta = 0°$

Figure S5 shows together the $R_{xx}$ and $R_{xy}$ curves of Sample 1 under perpendicular magnetic field at T=100 mK. While the longitudinal resistance indicates a clear minimum, only a faint feature is seen in the Hall resistance measurement. The gap associated to the skyrmion formation is such that the QH response is more developed in the $R_{xx}$ component.

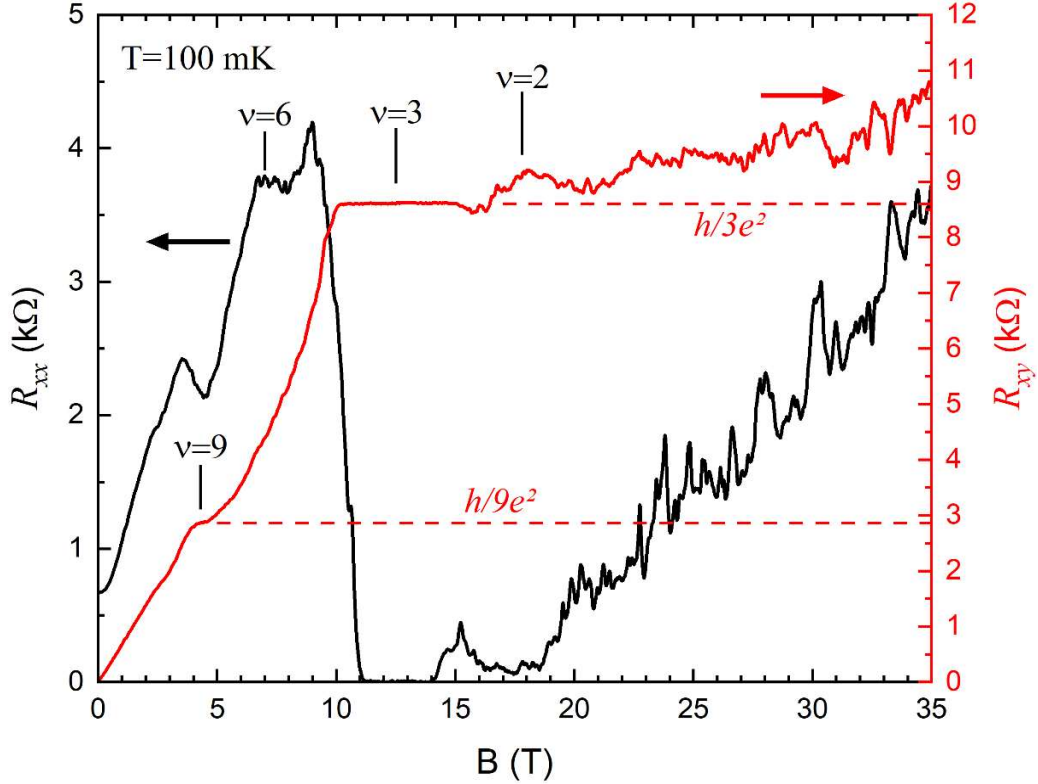

**FIG S5.** Longitudinal (in black) and transverse (in red) resistances of Sample 1 at T=100 mK and under perpendicular magnetic field. The observed quantized Hall resistances are indicated as well as the filling factors corresponding to observed $R_{xx}$ minima.